\documentclass[journal]{IEEEtran}
\ifCLASSINFOpdf
\else
\fi
\usepackage{cite}
\usepackage{graphicx} % for pdf, bitmapped graphics files
\usepackage{amsmath}
\usepackage{mathtools}
\usepackage{amssymb}
\usepackage{xcolor}
\usepackage{booktabs, tabularx}
\usepackage{fixdif}
\usepackage{empheq}
\usepackage{comment}
\usepackage{ragged2e}
\usepackage{bm}
\usepackage{url}
\usepackage{hyperref}
\usepackage{booktabs}
\usepackage[normalem]{ulem} % enables \sout
\usepackage{microtype}

\usepackage[caption=false,font=footnotesize]{subfig}

\usepackage{algorithm}
\usepackage{algpseudocode}
\usepackage{amsthm}
\theoremstyle{remark}

\AtBeginDocument{%
 
    \setlength{\abovecaptionskip}{4pt}

    \setlength{\textfloatsep}{8pt plus 2pt minus 2pt}
    \setlength{\dbltextfloatsep}{8pt plus 2pt minus 2pt}

    \setlength{\intextsep}{8pt plus 2pt minus 2pt}

    \setlength{\floatsep}{6pt plus 1pt minus 1pt}
    \setlength{\dblfloatsep}{6pt plus 1pt minus 1pt}
}

\allowdisplaybreaks[4]
\begin{document}
\title{Open-Source Multi-Fidelity EMT Modeling of Data Center Loads for Power System Studies}

\author{Xingyu~Zhao,~\IEEEmembership{Student~Member,~IEEE,~}Yingyi~Tang,~\IEEEmembership{Student~Member,~IEEE,~}and~Junbo~Zhao,~\IEEEmembership{Senior~Member,~IEEE}

\thanks{This work is partially supported by the U.S. National Science Foundation under grants 2328241 and 2550498.}% <-this % stops a space
\thanks{\textcolor{black}{X. Zhao, Y. Tang, and J. Zhao are with Thayer School of Engineering, Dartmouth College, Hanover, NH 03755 USA. (e-mail: junbo.zhao@dartmouth.edu)}}% <-this % stops a space

}
\markboth{}%
{Shell \MakeLowercase{\textit{et al.}}: Bare Demo of IEEEtran.cls for IEEE Journals}
\maketitle

\begin{abstract}
Characterizing data center–grid interactions under rapid load fluctuations requires electromagnetic transient (EMT) models that capture fast converter and control dynamics. This paper develops a modular multi-fidelity EMT model of data center loads in \texttt{PSCAD}. The modular design supports flexible assembly of subsystem models for different power supply configurations. The multi-fidelity design incorporates switching models, average-converter models, and reduced-order models to balance simulation accuracy and computational efficiency according to analysis needs. The model also provides a systematic controller tuning procedure and supports control mode switching. Comparative simulations demonstrate how this data center load model responds to both grid-side and load-side disturbances and assess the accuracy and computational efficiency of models with three fidelity levels. The model is publicly available on GitHub.\footnote{\url{https://github.com/xingyuz-phd/Generic-EMT-Modeling-of-Data-Center-Load}}
\end{abstract}

% Note that keywords are not normally used for peer-reviewed papers.
\IEEEpeerreviewmaketitle

\section{Introduction}
\IEEEPARstart{U}{nlike} conventional loads, large power-electronic loads such as AI data centers exhibit multi-timescale dynamics shaped by interacting control loops. Reported events in the United States include 14.7-Hz oscillations associated with a data center in Virginia~\cite{mishra2025oscillations}, 1-Hz forced oscillations at a data center in the Midwest~\cite{nerc2025largeloads}, and 23-Hz oscillations associated with a large electronic load in ERCOT~\cite{fan2026oscillations}. Beyond oscillations, a 3.8-GW data center load transfer to backup power following a July 2026 Dominion fault highlighted voltage ride-through concerns~\cite{pjm2026loadtransfer}. However, the mechanisms underlying these events are not fully understood, motivating the development of EMT models that capture fast transients and control interactions.

Existing dynamic models of data centers fall into two broad categories: phasor-domain and EMT-domain models. Phasor-domain models, including PERC1~\cite{weber2025perc1} and data center-specific representations~\cite{jimenez2026datacenter,gyang2026datacenter,lu2026calibration,zhao2026resonance}, offer computational efficiency for large power system studies but do not resolve instantaneous waveforms and fast electromagnetic transients. EMT models capture these dynamics through detailed representations of circuits and controls. Detailed models incorporating uninterruptible power supply (UPS) system, power supply units (PSUs), and cooling loads have been evaluated through simulation and hardware emulation~\cite{sun2022datacenter}. Reusable \texttt{PSCAD} libraries provide component and control models for studying voltage ride-through, operating mode transitions, and responses to computing load fluctuations~\cite{ross2026emt,shah2026emt,wang2026pscad}. However, interactions among interconnected converters complicate controller parameter tuning, highlighting the need for a systematic tuning method. Moreover, resolving switching transients and fast control dynamics can require very small simulation time steps, motivating reduced-order EMT models that preserve the dynamics of interest while improving computational efficiency for large power system studies.

\begin{figure}[!b]
    \centering
    \setlength{\abovecaptionskip}{0pt}
\includegraphics[width=\linewidth]{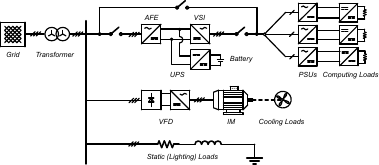}
\vspace{-10pt}
\caption{Schematic of the aggregated data center load model, comprising UPS-fed computing loads, VFD-driven cooling loads, and static lighting loads.}
    \label{fig:dc-layout}
\end{figure}

\begin{figure*}[!hp]
    \centering
    \setlength{\abovecaptionskip}{0pt}
\includegraphics[width=\textwidth]{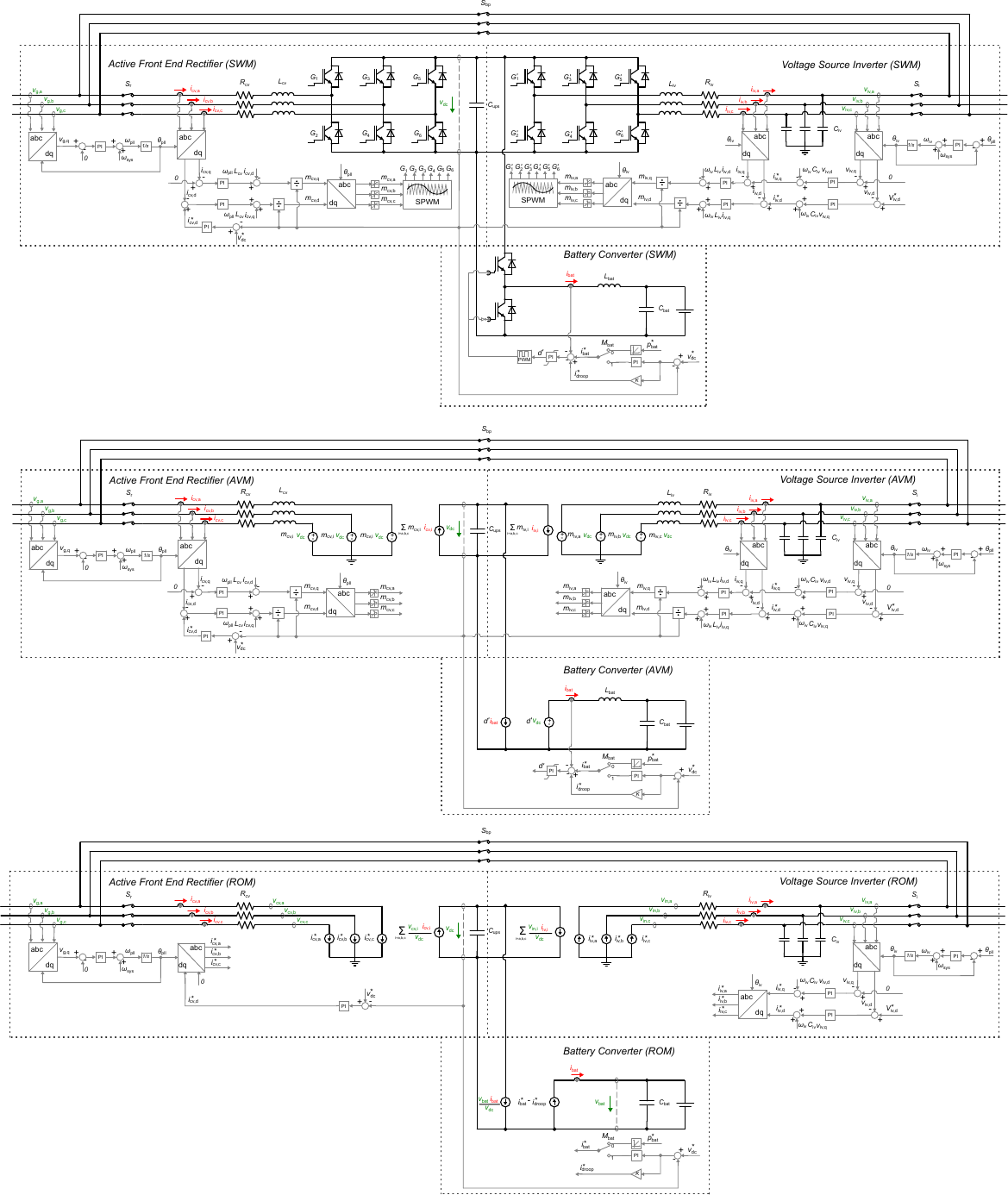}
\vspace{-10pt}
\caption{Circuit diagrams and control structures of the UPS switching model (top), average converter model (middle), and reduced-order model (bottom).}
    \label{fig:ups-models}
\end{figure*}

This paper develops an open-source EMT model of data center with selectable fidelity levels and systematic controller tuning based on prescribed closed-loop bandwidths and damping ratios. Simulation results illustrate the models’ responses to grid-side and load-side disturbances and demonstrate consistency among models with different levels of fidelity.

\section{Proposed Data Center EMT Model}
\subsection{Overview} 
Fig.~\ref{fig:dc-layout} illustrates the aggregated electrical layout of a data center, comprising computing, cooling, and static lighting loads. Computing loads are supplied by a UPS through PSUs. The UPS incorporates an active front-end (AFE) rectifier, a voltage-source inverter (VSI), and a battery energy storage system (BESS) for backup power. Cooling loads are modeled by a induction machine (IM) driven by a variable-frequency drive (VFD). Lighting loads are modeled as a static impedance. In practice, a data center may contain multiple UPS systems, with multiple PSUs connected to each output phase. Accordingly, the simulation model might be appropriately scaled to account for the number and power ratings of the UPS systems and PSUs.

\subsection{Uninterruptible Power Supply}
{\color{black} Fig.~\ref{fig:ups-models} shows the three UPS models developed in this work: the switching model (SWM), the average converter model (AVM), and the reduced-order model (ROM).}

\subsubsection{Mode Switching}
The UPS shown in Fig.~\ref{fig:ups-models} has three operating modes. In normal (double-conversion) mode, $S_{\mathrm{r}}$ and $S_{\mathrm{i}}$ are closed and $S_{\mathrm{bp}}$ is open, supplying grid power to the load through the AFE and VSI. When grid disturbances or outages prevent adequate supply, the UPS switches to battery backup mode, in which only $S_{\mathrm{i}}$ is closed and the BESS supplies the load through the battery converter and VSI. Bypass mode may be selected to facilitate UPS maintenance or improve efficiency when grid conditions are acceptable. In this mode, $S_{\mathrm{bp}}$ is closed and $S_{\mathrm{i}}$ is open, allowing direct grid supply; $S_{\mathrm{r}}$ may remain closed for battery charging or be opened to disconnect the AFE.

\subsubsection{Ceasing \& Reconnection Logic}
\begin{figure}[!b]
    \centering
    \setlength{\abovecaptionskip}{0pt}
    \includegraphics[width=0.85\linewidth]{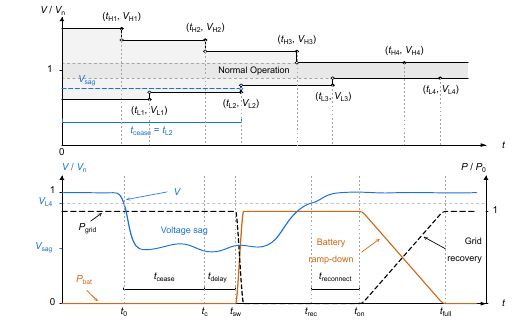}
    \vspace{-5pt}
    \caption{Ceasing and reconnection logic:
    VRT envelope (top) and voltage and power responses
    during UPS transfer and grid reconnection (bottom).}
    \label{fig:vrt-characteristics}
\end{figure}

Fig.~\ref{fig:vrt-characteristics} illustrates the voltage ride-through (VRT) envelope and the associated ceasing and reconnection logic. The upper panel defines the allowable voltage limits as a function of disturbance duration. In the proposed model, users can customize their VRT curves by specifying four $t\!-\!V$ coordinates for each boundary, \((t_{\mathrm{H}i}, V_{\mathrm{H}i})\) and \((t_{\mathrm{L}i}, V_{\mathrm{L}i})\), where \(i=1,\ldots,4\), using a sample-and-hold representation. The lower panel shows the response to a voltage sag. The VRT curve determines the permissible ride-through duration \(t_{\mathrm{cease}}\). If the sag persists, the UPS switches to battery backup mode after an additional delay \(t_{\mathrm{delay}}\). Once the voltage returns to the acceptable range and remains there for \(t_{\mathrm{reconnect}}\), the UPS transitions back to normal mode. During this transition, grid power supplied through the AFE and VSI gradually increases while power supplied by the BESS decreases, maintaining continuous load supply.

\subsubsection{Control Loops}
The AFE employs grid-following control in a $dq$ frame synchronized by PLL. Its outer DC-link voltage loop generates the $d$-axis current reference, while the $q$-axis reference is set to zero for unity power factor. The VSI forms the output voltage through cascaded PI loops: the voltage loop tracks the prescribed amplitude and generates current references, while the current loop produces modulation commands. Its internally generated phase is gradually aligned with the grid PLL angle through a low-bandwidth phase-alignment loop to reduce current transients during transfers between normal and bypass modes. For the battery converter, $M_{\mathrm{bat}}\!=\!1$ selects DC-link voltage control during battery backup, whereas $M_{\mathrm{bat}}\!=\!0$ selects constant-power control otherwise. A ramp-down rate limit on the power setpoint restores grid power draw gradually, mitigating power transients when normal operation resumes. Supplementary DC-voltage droop control adjusts the battery current reference to suppress UPS DC-link voltage fluctuations caused by workload variations and attenuate the propagation of load-side active-power oscillations to the grid during normal operation. To mitigate voltage and current transients caused by retained PI integral states during mode transitions, the outer-loop controllers are frozen and reset according to the switch states. When $S_{\mathrm{r}}$ is open, the AFE DC-link voltage PI output is held at zero, and its integral state is reset to zero. When $S_{\mathrm{i}}$ is open, the VSI outer voltage PI outputs are held constant, and their integral states are reset to zero. The inner current controllers of both converters remain active without freezing or resetting. The battery converter’s DC-link voltage PI integral state is reset whenever $M_{\mathrm{bat}}\!\neq\!1$, ensuring a zero initial integral state upon entry into battery backup mode.

\subsubsection{Model Reduction}

As shown in Fig.~\ref{fig:ups-models}, the AVM is derived from the SWM by replacing the semiconductor bridges and PWM generators with averaged controlled sources. Switching-frequency ripple is omitted, while passive-circuit dynamics and all control loops are retained. The ROM further assumes ideal current tracking and replaces the averaged converter stages with controlled current sources. Filter-inductor dynamics, inner current PI controllers, and modulation stages are removed, while DC-link and output-voltage dynamics, outer control loops, synchronization, and operating-mode logic are retained. Converter-port power balance is preserved through the DC-side current calculation.
\begin{figure}[!b]
    \centering
    \setlength{\abovecaptionskip}{0pt}
\includegraphics[width=\linewidth]{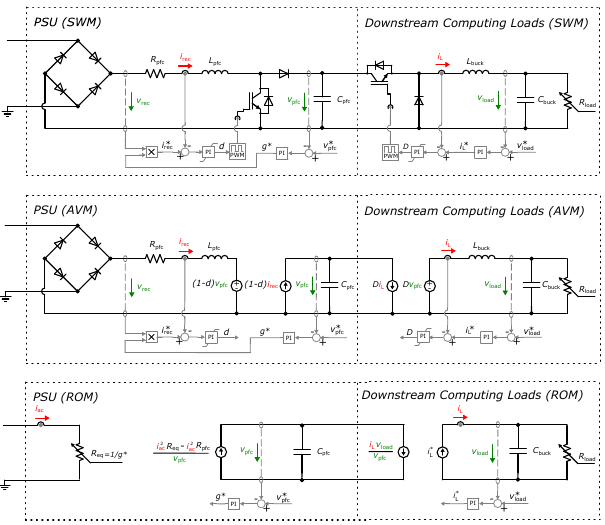}
\vspace{-15pt}
\caption{Circuit diagrams and control structures of the PSU and downstream IT load switching model (top), average-converter model (middle), and reduced-order model (bottom).}
    \label{fig:psu-models}
\end{figure}

\subsection{Power Supply Unit \& Downstream Computing Loads}

\subsubsection{Control Loops}
As shown in Fig.~\ref{fig:psu-models}, the PSU comprises a diode bridge followed by a boost-type PFC stage. The outer voltage PI loop regulates its output voltage $v_{\mathrm{pfc}}$ and generates an equivalent input conductance reference $g^{*}$. A multiplier forms the rectified-current reference, $i_{\mathrm{rec}}^{*}\!=\!g^{*}v_{\mathrm{rec}}$, and the inner current PI loop generates the boost duty ratio $d$. This shapes the AC input current to follow the supply-voltage waveform in phase, targeting unity power factor while maintaining the regulated DC voltage. The downstream computing loads are modeled by a voltage-regulated buck converter with a variable resistance $R_{\mathrm{load}}$ that emulates workload variations. Its outer voltage PI loop generates the inductor-current reference $i_L^{*}$, while the inner current PI loop determines the duty ratio $D$.

\begin{figure}[!b]
    \centering
    \setlength{\abovecaptionskip}{0pt}
\includegraphics[width=0.85\linewidth]{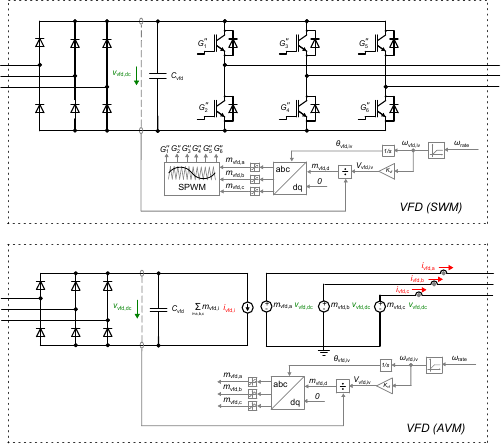}
 \vspace{-5pt}
\caption{Circuit diagrams and control structures of the VFD switching model (top) and average converter model (bottom).}
    \label{fig:vfd-models}
\end{figure}

\subsubsection{Model Reduction}

The AVM is derived from the SWM by replacing the switching stages and PWM generators with averaged controlled sources. Switching-frequency ripple is omitted, while LC circuit dynamics and both voltage and current control loops are retained. The ROM further assumes ideal current tracking, such that each inductor current instantaneously follows its reference. Under this assumption, the PFC consequently presents an equivalent resistive input impedance $R_{\mathrm{eq}}=1/g^{*}$ to the AC side with voltage $v_{\mathrm{ac}}$, while the buck current source follows $i_L^{*}$. The capacitor dynamics of $C_{\mathrm{pfc}}$ and $C_{\mathrm{buck}}$ and the outer voltage PI loops are retained. Converter-port power balance is preserved through the coupling-current calculation.

\subsection{Variable Frequency Drive \& Cooling Loads}

As shown in Fig.~\ref{fig:vfd-models}, the VFD implements open-loop
scalar V/f control~\cite{munoz1998vf}, with the voltage amplitude proportional to
the commanded frequency.  The cooling loads are represented using \texttt{PSCAD}'s built-in induction IM model operating in torque-input mode. The mechanical load is modeled using a quadratic torque--speed
relationship, $T_{\mathrm{L}}=K_{\mathrm{L}}\omega_{\mathrm{m}}^{2}$,
consistent with centrifugal-pump characteristics~\cite{singh2018pump}.
The coefficient is determined as
$K_{\mathrm{L}}=p_{\mathrm{shaft}}/\omega_{\mathrm{rate}}^{3}$,
where $p_{\mathrm{shaft}}$ is the shaft power at the rated speed
$\omega_{\mathrm{rate}}$.

\subsection{Control Parameter Tuning}

The PI gains are derived by approximating the plant as a
first-order system and matching the resulting nominal
second-order closed-loop characteristic polynomial to
$s^2\!+\!2\zeta\omega_n s\!+\!\omega_n^2$.
Current-loop tuning uses the inductor dynamics, whereas
voltage-loop tuning uses the capacitor dynamics with a
lower bandwidth and a fast-current-tracking approximation.
Taking the VSI in Fig.~\ref{fig:ups-models} as an example,
with $dq$ decoupling in voltage and current control,
the PI gains are 
$K_{p,\ell}\!=\!2\zeta_\ell\omega_{n,\ell}L_{\mathrm{iv}}/\omega_b\!-\!R_{\mathrm{iv}}$,
$K_{i,\ell}\!=\!\omega_{n,\ell}^{2}L_{\mathrm{iv}}/\omega_b$,
$K_{p,v}\!=\!2\zeta_v\omega_{n,v}C_{\mathrm{iv}}/\omega_b$, and
$K_{i,v}\!=\!\omega_{n,v}^{2}C_{\mathrm{iv}}/\omega_b$.
Here, subscripts $\ell$ and $v$ denote the current and voltage loops, respectively.
The natural frequency $\omega_n\!=\!2\pi f_{\mathrm{bw}}$
is used as the bandwidth-setting parameter for PI tuning,
while $\zeta$ specifies the damping ratio.
The parameters $L_{\mathrm{iv}}$, $R_{\mathrm{iv}}$,
and $C_{\mathrm{iv}}$ are in p.u., and $\omega_b$ is the
base angular frequency.

\section{Simulation Results}
\begin{table}[!b]
\caption{Default Parameters of UPS-Fed Computing Load}
\label{tab:ups_computing_parameters}
\centering
\footnotesize
\setlength{\tabcolsep}{1.5pt}
\setlength{\aboverulesep}{0.15ex}
\setlength{\belowrulesep}{0.15ex}
\renewcommand{\arraystretch}{0.92}

\begin{tabular*}{\columnwidth}{@{\extracolsep{\fill}}llr@{}}
\toprule
\multicolumn{1}{@{}l}{Symbol(s)}
& \multicolumn{1}{c}{Description}
& \multicolumn{1}{@{}r}{Value(s)} \\
\midrule

$S_{\mathrm{ups,b}}$
& UPS power base & $1.5$ MVA \\
$S_{\mathrm{psu,b}}$
& PSU \& buck power base & $0.05$ MVA \\
$V_{\mathrm{ac,b}}$
& UPS \& PSU ac voltage base (LL-RMS) & $0.48$ kV \\
$V_{\mathrm{dc,b}}$
& UPS dc-link nominal voltage & $0.8$ kV \\
$V_{\mathrm{pfc,b}}$
& PFC dc output voltage base & $0.8$ kV \\
$V_{\mathrm{bat,b}}$
& Battery nominal voltage/base & $0.4$ kV \\
$V_{\mathrm{load,b}}$
& Buck dc output (load) voltage base & $0.048$ kV \\
$\omega_b$
& Base frequency & $2\pi\times60$ Hz \\

$R_{\mathrm{cv}}$
& AFE ac filter resistance & $0.003$ p.u. \\
$L_{\mathrm{cv}}$
& AFE ac filter inductance & $0.2$ p.u. \\
$C_{\mathrm{ups}}$
& UPS dc-link capacitance & $10$ p.u. \\
$f_{\mathrm{bw},\ell},\,\zeta_\ell$
& AFE current loop & $400$ Hz, $0.707$ \\
$f_{\mathrm{bw},v},\,\zeta_v$
& AFE dc-voltage loop & $20$ Hz, $0.707$ \\
$f_{\mathrm{bw},\mathrm{pll}},\,\zeta_{\mathrm{pll}}$
& PLL & $15$ Hz, $1$ \\

% Battery parameters
$Q_{\mathrm{bat,r}}$
& Battery rated capacity & $0.65$ kAh \\
$\delta_{\mathrm{cap,1h}}$ &Capacity loss/hour at nominal current 
& $300\%$ \\
$Q_{\mathrm{bat,n}}$
& Battery nominal capacity & $0.95$ p.u. \\
$\Delta V_{\mathrm{bat,R}}$
& Battery resistive voltage drop & $0.005$ p.u. \\
$V_{\mathrm{bat,exp}}$
& Voltage at exponential point & $1.03$ p.u. \\
$Q_{\mathrm{bat,exp}}$
& Capacity at exponential point & $0.4$ p.u. \\
$V_{\mathrm{bat,full}}$
& Fully charged battery voltage & $1.15$ p.u. \\

% Battery converter parameters
$L_{\mathrm{bat}}$
& Battery converter inductance & $0.2$ p.u. \\
$C_{\mathrm{bat}}$
& Battery converter capacitance & $0.5$ p.u. \\
$f_{\mathrm{bw},\ell},\,\zeta_\ell$
& Battery converter current loop & $500$ Hz, $0.707$ \\
$f_{\mathrm{bw},v},\,\zeta_v$
& Backup voltage loop & $20$ Hz, $1$ \\
$p_{\mathrm{bat}}^{*}$
& Battery power setpoint & $0$ p.u. \\
$K$
& Battery dc-voltage droop gain & $0$ \\
$r_{\mathrm{lim}}$
& Power ramp-down rate limit & $0.5$ p.u./s \\

$R_{\mathrm{iv}}$
& VSI filter resistance & $0.003$ p.u. \\
$L_{\mathrm{iv}}$
& VSI filter inductance & $0.05$ p.u. \\
$C_{\mathrm{iv}}$
& VSI filter capacitance & $1$ p.u. \\
$f_{\mathrm{bw},\ell},\,\zeta_\ell$
& VSI current loop & $500$ Hz, $0.707$ \\
$f_{\mathrm{bw},v},\,\zeta_v$
& VSI voltage loop & $100$ Hz, $1$ \\
$f_{\mathrm{bw},\theta},\,\zeta_\theta$
& VSI phase alignment loop & $0.5$ Hz, $1$ \\

$R_{\mathrm{pfc}}$
& PFC filter resistance & $0.003$ p.u. \\
$L_{\mathrm{pfc}}$
& PFC filter inductance & $0.1$ p.u. \\
$C_{\mathrm{pfc}}$
& PFC dc-link capacitance & $10$ p.u. \\
$f_{\mathrm{bw},\ell},\,\zeta_\ell$
& PFC current loop & $5000$ Hz, $0.707$ \\
$f_{\mathrm{bw},v},\,\zeta_v$
& PFC voltage loop & $10$ Hz, $1$ \\

$L_{\mathrm{buck}}$
& Buck inductance & $5$ p.u. \\
$C_{\mathrm{buck}}$
& Buck output capacitance & $2.5$ p.u. \\
$f_{\mathrm{bw},\ell},\,\zeta_\ell$
& Buck current loop & $500$ Hz, $0.707$ \\
$f_{\mathrm{bw},v},\,\zeta_v$
& Buck voltage loop & $20$ Hz, $1$ \\

$t_{\mathrm{delay}},\,t_{\mathrm{reconnect}}$
& UPS ceasing \& reconnection delay & $0.2$\,s, $0.5$\,s \\
$f_{\mathrm{sw}}$
& Switching frequency & $10$ kHz \\

\bottomrule
\end{tabular*}

\par\vspace{1pt}
\end{table}

\subsection{UPS-Fed Computing Loads}

The SWM, AVM, and ROM of the UPS-fed computing loads in
Fig.~\ref{fig:dc-layout} are compared under grid-side and
workload-side disturbances. Each UPS supplies three
single-phase PSUs, each scaled by a factor of 10, and the
complete UPS-fed branch is scaled by a factor of 30,
giving an aggregate power base of $45\,\mathrm{MVA}$.
The grid is represented by an ideal $0.48\,\mathrm{kV}$
voltage source. The battery utilizes the Shepherd model. Model parameters are listed in
Table~\ref{tab:ups_computing_parameters}.
The SWM uses a simulation time step of $2\,\mu\mathrm{s}$,
whereas the AVM and ROM use $10\,\mu\mathrm{s}$. 
During the initialization stage ($0 \leq t < 2\,\mathrm{s}$),
the IT workload is initialized to $0.1$~p.u. and ramped
to $0.6$~p.u. A grid-side voltage sag is applied at
$t=2\,\mathrm{s}$, with recovery beginning at
$t=2.5\,\mathrm{s}$. Periodic square-wave workload
variations are introduced from $t=6\,\mathrm{s}$,
representing demand fluctuations associated with
AI training tasks.

\begin{figure}[!t]
    \centering
    \setlength{\abovecaptionskip}{2pt}

    \includegraphics[width=\linewidth]{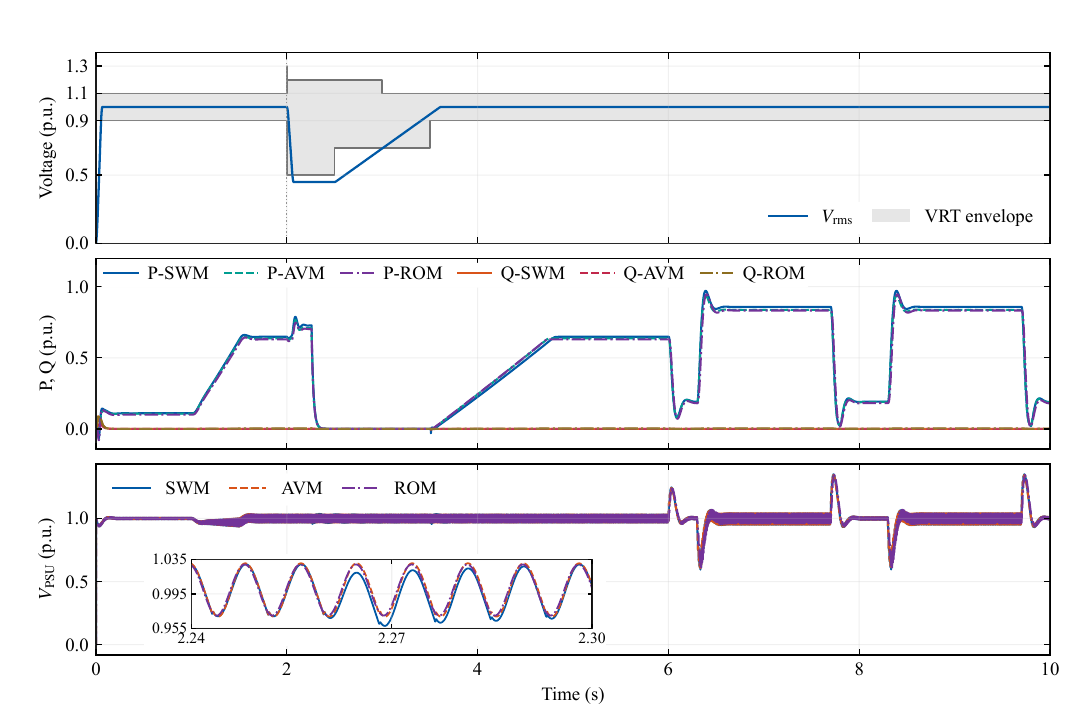}
        \vspace{-22pt}
    \caption{SWM, AVM, and ROM responses to grid and workload
disturbances. From top to bottom: grid voltage and VRT
envelope, UPS grid-side active and reactive power, and PSU dc voltage.}
    \label{fig:model_comparison}
\end{figure}

Fig.~\ref{fig:model_comparison} illustrates the response of the model to a voltage sag and subsequent workload variations. When the measured RMS voltage violates the VRT envelope, the UPS transfers to battery backup after a delay while maintaining the PSU dc output voltage near its nominal value. Once the voltage returns to the normal operating region, grid power draw is gradually restored subject to the specified power ramp limit. The three models show close agreement in operating-mode transitions and responses to workload variations. The SWM predicts slightly higher steady-state active-power demand than the AVM and ROM. This difference may be partly attributable to additional resistive losses associated with modulation-induced current ripple and waveform distortion. The ROM achieves twice the simulation speed of the AVM under the same simulation setup.

\begin{table}[!b]
\caption{Default Parameters of VFD-Driven Cooling Load}
\label{tab:cooling_parameters}
\centering
\footnotesize
\setlength{\tabcolsep}{1.5pt}
\setlength{\aboverulesep}{0.15ex}
\setlength{\belowrulesep}{0.15ex}
\renewcommand{\arraystretch}{0.92}

\begin{tabular*}{\columnwidth}{@{\extracolsep{\fill}}llr@{}}
\toprule
\multicolumn{1}{@{}l}{Symbol(s)}
& \multicolumn{1}{c}{Description}
& \multicolumn{1}{@{}r}{Value(s)} \\
\midrule
$S_{\mathrm{vfd,b}}$
& VFD \& IM power base & $0.5$ MVA \\
$V_{\mathrm{ac,b}}$
& Grid-side ac voltage base (LL-RMS) & $0.48$ kV \\
$V_{\mathrm{im,b}}$
& IM rated voltage (LL-RMS) & $0.32$ kV \\
$f_{\mathrm{im,b}}$
& IM base frequency & $40$ Hz \\
$C_{\mathrm{vfd}}$
& VFD dc-link capacitance & $10$ p.u. \\
$\omega_{\mathrm{rate}}$
& Motor speed setpoint & $0.98$ p.u. \\
$V_{\mathrm{vfd,max}}$
& Maximum converter voltage & $1.2$ p.u. \\
$R_{\mathrm{s}},\,R_{\mathrm{r}}$
& Stator \& rotor resistance & $0.015,\,0.012$ p.u. \\
$L_{\mathrm{m}}$
& IM magnetizing inductance & $3.0$ p.u. \\
$L_{\mathrm{ls}}, L_{\mathrm{lr}}$
& Stator \& Rotor leakage inductance & $0.08, 0.08$ p.u. \\
$H$
& IM inertia constant & $0.5$ s \\
$D_{\mathrm{m}}$
& Mechanical damping coefficient & $0.01$ p.u. \\
$p_{\mathrm{shaft}}$ & Shaft power &  $0.5$ p.u.\\ 
$f_{\mathrm{sw}}$
& Switching frequency  & 1 kHz \\ 
\bottomrule
\end{tabular*}

\par\vspace{1pt}
\end{table}

\subsection{VFD-Driven Cooling Load}

\begin{figure}[!t]
    \centering
    \setlength{\abovecaptionskip}{2pt}

    \includegraphics[width=\linewidth]{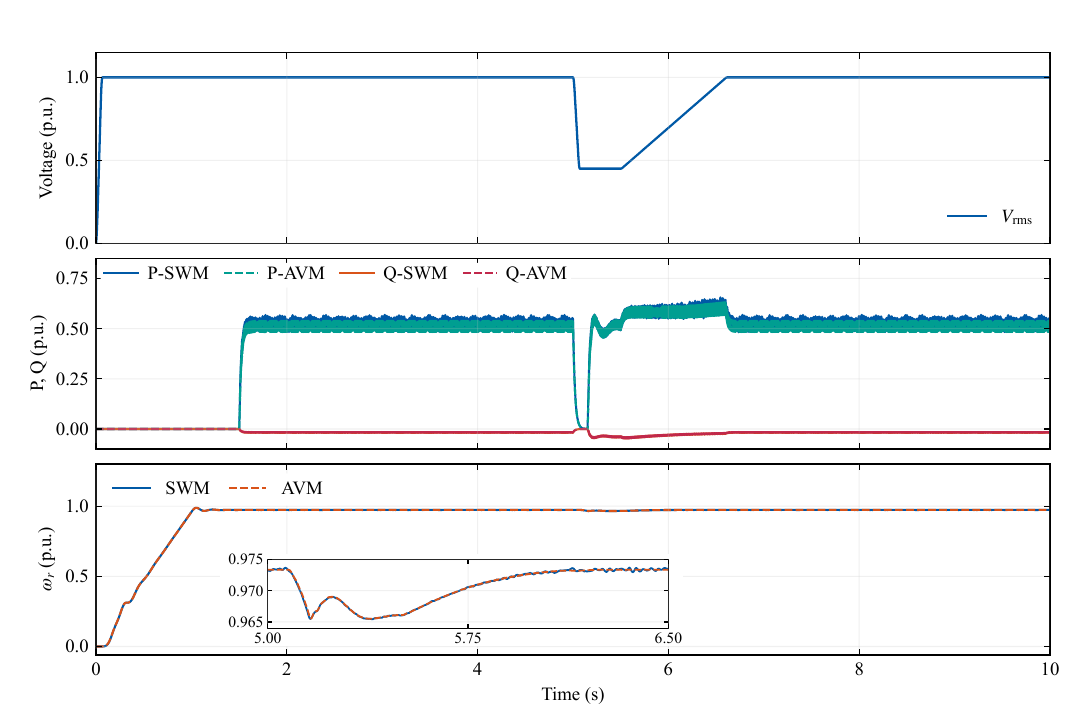}
        \vspace{-22pt}
\caption{SWM and AVM responses of the cooling load.
From top to bottom: grid voltage, VFD active and
reactive power, and induction-motor speed.}
    \label{fig:Cooling_comparison}
\end{figure}

The SWM and AVM of the VFD-driven cooling loads are compared
under a grid-side voltage disturbance, with model parameters
listed in Table~\ref{tab:cooling_parameters}. Both models uses a simulation time step of 5 $\mu s$. Following initialization over $0 \leq t < 5\,\mathrm{s}$,
the same voltage-sag profile used in the UPS-fed computing-load
case is applied at $t=5\,\mathrm{s}$, with recovery beginning
at $t=5.5\,\mathrm{s}$. Fig.~\ref{fig:Cooling_comparison} shows close agreement
between the SWM and AVM in active-power, reactive-power,
and motor-speed responses during the voltage sag and
subsequent recovery.

\section{Conclusion}
This paper presents an open-source, multi-fidelity EMT model of data center loads.
Simulations demonstrate consistent dynamic responses across different fidelity levels.
Future work will focus on experimental validation and large-scale studies.

\end{document}